\documentstyle[12pt]{article}
\begin{document}
\title{\vspace{-15mm}
       {\normalsize \hfill
       \begin{tabbing}
       \`\begin{tabular}{l}
	DESY 94--237 \\
	 December 1994 \\
	 hep--th/9412162 \\
	\end{tabular}
       \end{tabbing} }
       \vspace{8mm}
\setcounter{footnote}{1}
The 10-D chiral null model and the relation \\
to 4-D string solutions
}

\renewcommand{\thefootnote}{\fnsymbol{footnote}}

\author{
Klaus Behrndt\thanks{e-mail: behrndt@ifh.de, Work
 supported in part by a grant of the DAAD} \\ {\normalsize \em
 DESY-Institut f\"ur Hochenergiephysik, Zeuthen} \\
{\normalsize \em Platanenallee 6, 15735 Zeuthen, Germany}
}
\date{}
\maketitle
\renewcommand{\arraystretch}{2.0}
\renewcommand{\thefootnote}{\alph{footnote}}
\newcommand{\be}{\begin{equation} \label{\theequation}}
\newcommand{\ee}{\end{equation}}
\newcommand{\ba}{\begin{array}}
\newcommand{\ea}{\end{array}}
\newcommand{\vsf}{\vspace{5mm}}
\newcommand{\NP}[3]{{\em Nucl. Phys.}{ \bf B#1#2#3}}
\newcommand{\PRD}[2]{{\em Phys. Rev.}{ \bf D#1#2}}
\newcommand{\PRL}[2]{{\em Phys. Rev. Lett.}{ \bf #1#2}}
\newcommand{\MPLA}[1]{{\em Mod. Phys. Lett.}{ \bf A#1}}
\newcommand{\PL}[3]{{\em Phys. Lett.}{ \bf B#1#2#3}}
\newcommand{\marpar}{\marginpar[!!!]{!!!}}

\begin{abstract} \noindent
  The chiral null model is a generalization of the fundamental string
  and gravitational wave background. It is an example of a conformally
  invariant model in all orders in $\alpha'$ and has unbroken
  supersymmetries.  In a Kaluza--Klein approach we start in 10
  dimensions and reduce the model down to 4 dimensions without making
  any restrictions. The 4-D field content is given by the metric,
  torsion, dilaton, a moduli field and 6 gauge fields. This model is
  self-dual and near the singularities asymptotically free.  The
  relation to known IWP, Taub-NUT and rotating black hole solutions is
  discussed.
\end{abstract}

\noindent
In order to make statements about ``stringy'' modification of the
point particle physics it is necessary to find solutions, which solve
the equation of motion also in higher orders in $\alpha'$. Especially,
if one wants to investigate space time singularities in the string
theory one needs solutions in all orders. A couple of exact solutions
could be found in recent years. There are mainly two classes of
solutions. One is given by various combinations of (gauged) WZW
theories and the other one contains solutions for which the $\alpha'$
corrections vanish identically. The chiral null model \cite{ho/ts1},
e.g.\ , belongs to last class of exact solutions.  Special limits of
this model are the gravitational plane waves and the fundamental
string background. Embedded in N=1, D=10 supergravity it has been
shown that these models admit unbroken supersymmetries \cite{ka}
and that an extension to (0,1) world sheet supersymmetry is possible
\cite{ho/ts1}.

In this paper we are going to discuss the dimensional reduction of the
chiral null model. First, we summarize some general properties of this
model. After a dimensional reduction of this model from 10 to 4
dimensions we discuss relations to other known 4-D string background.

\vspace{3mm}

\noindent
The model is given by
\be
\ba{l}
ds^2 = 2 F(x) du \,[ dv - \frac{1}{2} K(x) du + \omega_i(x) dx^i ] -
dx^i dx^i \\
\hat{B} = 2 F(x) du \wedge [ dv + \omega_i(x) dx^i] \qquad ,
     \qquad \hat{\phi} = \hat{\phi}(x)
\ea
\ee
where the space-time is spanned by the coordinates $\{u , v , x^i \}$
and we are assuming that the fields does not depend on $u$ and $v$.
The quantities with hat are higher dimensional objects in contrast
to the dimensional reduced quantities we discuss below.
The corresponding world sheet Lagrangian is given by
\be
L = 4 F(x) \partial u \, [ \, \bar{\partial} v - K(x)\bar{\partial} u +
   \omega_i(x) \bar{\partial} x^i \,] - \partial x^i \bar{\partial} x^i +
   \alpha' R^{(2)} \hat{\phi}(x)
\ee
A crucial property of this model is the chiral symmetry: $ v
\rightarrow v + h(\tau + \sigma)$. In the target space this world sheet
symmetry is manifest in the existence of a null killing vector: $k_{\mu} =
\{0,1,0,..,0\}$ ($k_v =1$). This symmetry, especially the chiral
coupling of the vector field $\omega_i$, ensures the vanishing of all
higher $\alpha'$ corrections in the renormalization group $\beta$
functions.  After integrating out $u$ and $v$ it is possible to show
that for the renormalization only tadpole diagrams are relevant (see
\cite{ho/ts1}). The chiral structure of the Lagrangian makes it
impossible to construct other (non-tadpole) divergent diagrams.  Thus,
the conformal invariance conditions are given by the lowest order in
$\alpha'$ and if we drop a linear dilaton part we have the equations
\be
-\partial^2 K(x) \, = \, -\partial^2 F^{-1}(x) \, = \, 0
\qquad , \qquad -\partial^i F_{ij} \, = \, 0 \qquad \mbox{and} \qquad
 e^{2 \hat{\phi}} \sim F(x) \ .
\ee
($F_{ij}=\partial_{[i} \omega_{j]}$). Furthermore, since the model
does not depend on $u$ and $v$ we can dualize any non-null direction
in the $(u,v)$ plane, e.g.\ transforming $v = \hat{v} + c \,u$ and
dualizing $u$ yields
\be
F' = \left(c - \frac{1}{2} K\right)^{-1} \quad , \quad K' = F^{-1} - c \quad ,
\quad \omega'_i = \omega_i
\quad , \quad e^{-2 \hat{\phi}'} = e^{-2 \hat{\phi}} F \,
 \left(c - \frac{1}{2} K\right)= F'^{-1}
\ee
Using the fact that $K$ and $F^{-1}\sim e^{-2 \hat{\phi}}$ are harmonic
functions we see that the duality transformation changes only
the parameter of the solution, i.e.\ the model is explicitly self-dual.

There are two models of special interest. One are the plane fronted
waves or the $K$-model ($F=1$ or $\phi=\phi_0$) with the Lagrangian
\be
L_K = 4 \partial u \left[ \bar{\partial} v - K(x) \bar{\partial} u
    + \omega_i \bar{\partial} x^i \right] - \partial x^i
        \bar{\partial} x^i \ .
\ee
For this model the killing vector becomes even a covariant constant null
vector. The other model is given by the fundamental string model or
$F$-model ($K=0$)
\be
L_F = 4 e^{2 \hat{\phi}} \partial u \left[ \bar{\partial} v +
  \omega_i \bar{\partial} x^i \right] -
  \partial x^i \bar{\partial} x^i  + \alpha' R^{(2)} \hat{\phi} \ .
\ee
Here, we have a second chiral current corresponding
to a shift in $u$: $u \rightarrow u + f(\tau - \sigma)$. Consequently,
in the target space we have two null killing vectors: $k_{\mu}^{(1)} =
\partial_{\mu} u$ and $k_{\mu}^{(2)} = \partial_{\mu} v$. From (\ref{4})
it follows that both models are dual to each other (setting $c=1$).

\vspace{3mm}

\noindent
It is possible to show that one can embed this model in N=1, D=10
supergravity with unbroken supersymmetries \cite{ka}. On the other
side it is possible to extend this model to a (0,1) world sheet
supersymmetry \cite{ho/ts1}.  Therefore, it is reasonable to consider
this model from the point of view of 10-D superstring theory.  Since
there are no higher $\alpha'$ corrections the complete effective
action (up to non-perturbative and higher genus contribution) is given by
\be
S_{10} = \int d^{10} X \sqrt{\hat{G}} e^{- 2 \hat{\phi}} \left[ \; R +
        4 (\partial \hat{\phi})^2 - \frac{1}{12} \hat{H}^2 \; \right]
\ee
where: $X^M = \{v,x^1,...,x^8,u\}$, $\hat{H} = d\hat{B}$ and $\hat{B}$
is the 10-d antisymmetric tensor. Our aim is now to reduce this 10-D
theory down to 4 dimensions and to discuss the resulting background.
Since the reduction procedure preserves the supersymmetry also the 4-D
background has unbroken supersymmetries (corresponding to N=4).
Assuming that the theory does not depend on 6 coordinates and that
this internal space is compact we can integrate over these coordinates
and get the 4-D theory. This is more or less standard in
string compactification (see e.g.\ in \cite{ma/sc}). On the other side
if one admits a dependence on the internal coordinates, one can make a
Fourier expansion in the internal coordinates. After reduction one
gets then states with masses corresponding to the inverse
compactification scales (see e.g.~\cite{du}). In this philosophy we
are in the massless sector.

{}From (\ref{1}) we find for the 10-D metric and antisymmetric tensor
\be
\hat{G}_{MN} =  \left( \ba{cc|cc}
0 & 0 & 0  & F \\[-4mm]
0 & -\delta_{ij} & 0 & F \,\omega_i \\ \hline
0 & 0 & -\delta_{mn} & F\,\omega_m  \\
F & F\,\omega_i  &F\,\omega_m  & - F\,K \ea \right) \quad , \quad
\hat{B}_{MN} =  \left( \ba{cc|cc}
0 & 0 & 0  & F \\[-4mm]
0 & 0 & 0 & F\,\omega_i \\ \hline
0 & 0 & 0 & F\,\omega_m  \\
-F & -F\,\omega_i  &-F\,\omega_m  & 0 \ea \right)
\ee
where the first column corresponds to $v$ (becomes the time later);
$i,j = 1,2,3$ (spatial coordinates); $m,n = 4,5,...8$ and the last
column $u$ are internal coordinates. Following now the standard
procedure for dimensional reduction (see e.g.\ \cite{ma/sc,ch}) we write
the 10-Bein as
\be
e_M^{\ \hat{N}} = \left( \ba{c|c}
        e_{\mu}^{\ \hat{\mu}} & A_{\mu}^{\ r} E_r^{\ \hat{s}} \\ \hline
         0         &   E_r^{\ \hat{s}} \ea \right) \ .
\ee
The 4-D space-time metric is given by $g_{\mu\nu} =
e_{\mu}^{\hat{\mu}} e_{\nu}^{\hat{\nu}} \eta_{\hat{\mu} \hat{\nu}}$
and the internal metric is $G_{rs} =
E_{r}^{\hat{r}} E_{s }^{\hat{s }} \delta_{\hat{r } \hat{s }}$ ($r,s = 4,..
9$).  This form of the 10-Bein has the advantage that the
determinant and thus the volume measure factorizes and we can absorb
the internal part in the dilaton
\be
\sqrt{|\hat{G}_{MN}|} \, e^{-2 \hat{\phi}} = |e_M^{\ N}|
\, e^{-2 \hat{\phi}} =
\sqrt{|g_{\mu\nu}|} \sqrt{|G_{rs}|} \, e^{-2 \hat{\phi}} =
\sqrt{|g_{\mu\nu}|} e^{-2 \phi}
\ee
with the 4-D dilaton $\phi$ defined by
\be
\sqrt{|G_{rs}|} \, e^{-2 \hat{\phi}} = e^{-2 \phi}  \ .
\ee
In terms of  the 10-Bein we get for the 10-D metric
\be
\hat{G}_{MN} = \left( \ba{c|c}
  g_{\mu\nu} + A^{~r}_{\mu} A_{\nu r} &
  A_{\mu r}\\ \hline
  A_{\mu s} & G_{rs} \ea \right) \ .
\ee
Comparing (\ref{12}) with (\ref{8}) we find for the internal metric
and the gauge field part
\be
G_{rs} =  \left(\ba{c|c} -  \delta_{mn} & F\,\omega_m \\[2mm] \hline
           F\,\omega_n  & - F K \ea \right)  \qquad , \qquad A_{\mu r} =
       \big( \  0 \  | \, A_{\mu} \, \big)
\ee
with $A_{\mu} = F\, \{1, \omega_i \} $ and inserting
this  metric we find for the 4-D dilaton (\ref{11})
\be
 e^{-2 \phi} =  e^{-2 \hat{\phi}}  \sqrt{F K - F^2\,|\omega_m|^2} =
\sqrt{F^{-1} K - |\omega_m|^2}
\ee
where $|\omega_m|^2 = \omega_4^2 + \omega_5^2 + .. $ . We assume in the
following that $F^{-1} K > |\omega_m|^2$. This means that all
eigenvalues of the internal metric are negative\footnote{The
eigenvalues of $G_{rs}$ are: $\left\{-1,-1,-1,-1,-\frac{1}{2}\left((1
+ F K) \pm \sqrt{(1 - F K)^2 + 4 F^2 |\omega_m|^2}\right)\right\}$.} ,
i.e.\ there is no timelike compactified coordinate. In addition, we
find for the 4-D metric
\be
ds^2 = \frac{1}{F^{-1} K - |\omega_m|^2} (dt^2 + \omega_i dx^i)^2 -
  dx^i dx^i =  e^{4 \phi} (dt^2 + \omega_i dx^i)^2 - dx^i dx^i .
\ee
Before we are going to discuss the metric let us derive the other 4-D
fields.  The 4-D antisymmetric tensor can be calculated in terms of
the 10-D components $\hat{B}_{\mu\nu}$ \cite{ma/sc} and vanishes in
our case
\be
B_{\mu\nu} = \hat{B}_{\mu\nu} - A_{\mu}^{\ r} \hat{B}_{rs}
 A_{\nu}^{\ s} = 0
\ee
In principle there is one further term, which however, vanishes in our
case too because the gauge fields from the metric and antisymmetric
tensor are equal. But nevertheless the 4-D torsion is non-vanishing
and is proportional to the Chern-Simons term of the gauge field
$A_{\mu}=F \{1,\omega\}$
\be
H_{\mu\nu\rho} = - A_{[\mu}^{\ \ r} \partial_{\nu} A_{\rho] s} =
         - e^{4 \phi}\, F^{-2}\, A_{[\mu} \partial_{\nu} A_{\rho]} \ .
\ee
The scalar field content is given by the dilaton and a modulus field.
The 4-D dilaton is given by (\ref{14}).  The easiest way to see that
there is only one modulus field is to perform the dimensional
reduction in two steps. First we reduce the coordinates $x^5 ... x^8$
and then the $u$ coordinate (last column in (\ref{8})). During the
first reduction the corresponding internal metric is flat and
therefore no modulus field appears. But the resulting 5-D metric has a
non-constant (5,5) component corresponding to the following modulus
field $\sigma$
\be
e^{\sigma} =  \sqrt{|G_{rs}|} = \sqrt{F K - F^{2} |\omega_m|^2} =
         e^{2 ( \hat{\phi} - \phi )} \ .
\ee
Following these two reductions it is clear that the geometry of the
internal space is the direct product of a constant 5-D torus
(corresponding to the flat internal metric of the first reduction) and
a circle with non-constant radius given by the moduli field $\sigma$.

Finally, we have to discuss the gauge field content. Gauge fields
appear in the Kaluza--Klein procedure as non-diagonal components of
the metric and the antisymmetric tensor. The gauge fields coming from
the metric is in principle given by (\ref{13}). But there is a
subtlety.  Investigating the gauge field transformation one realizes
that the basic gauge field has to have an upper internal index. The
reason is, that gauge transformation are generated by local
translations in the internal coordinates which have an upper index.
Rising the index we get
\be
A_{\mu}^{\ r} = A_{\mu s} G^{sr} = \frac{-1}{K - F\,|\omega_m|^2}
\left( \ \omega^n \,A_{\mu} \ | \ F^{-1}\,A_{\mu} \ \right)
\ee
($\omega^n = \omega_n$). On the other side, the gauge fields coming
from the antisymmetric tensor have a lower internal index. The
corresponding gauge transformation is part of the antisymmetric tensor
gauge symmetry. But here, to get the right gauge field coupling in the
effective action we have to add an additional term \cite{ma/sc} and
obtain
\be
B_{\mu r} = \hat{B}_{\mu r} + \hat{B}_{rs} A_{\mu}^{\ s} =
\frac{1}{K - F\,|\omega_m|^2} \left(\ \omega_m \,A_{\mu} \
\big| \  K\,A_{\mu}\ \right) \ .
\ee

\vspace{3mm}

\noindent
So far we have discussed the 4-D fields as function of the 10-D
quantities. Now, we have to solve the equations of motion and
to discuss the concrete solution. Assuming that all fields depend
only on the three spatial coordinates we get from (\ref{3})
\be
-\partial^2 K(x) \, = \, -\partial^2 F^{-1}(x) \, = -\partial^2 \omega_m(x)
\, = \, 0
\qquad \mbox{and} \qquad -\partial^{i} F_{ij} \, = \, 0 \ .
\ee
For the field strength $F_{ij} = \partial_{[i} \omega_{j]}$ we have two
solutions
\be
F_{ij} = \epsilon_{ijk} \partial^k \, a \qquad \mbox{or} \qquad F_{ij}
= const.
\ee
In the second case we have an uniform magnetic field and under certain
assumptions the target space is parallizable and the model corresponds
to a product of a non-semisimple WZW model and a free spatial
direction. The corresponding 4-D space time is not asymptotically
flat.  Let us ignore this case here (see \cite{ru/ts}).  The
first case corresponds to the known monopole background, which is
determined by a further harmonic scalar field $a$. Inserting this
gauge field in (\ref{17}) we find for the torsion
\be
\epsilon^{\lambda\mu\nu\rho} H_{\mu\nu\rho} =
    e^{2 \phi} \, \partial^{\lambda} \, a
\ee
where $\epsilon^{\lambda\mu\nu\rho}$ is the covariant epsilon tensor
in the string frame.  Therefore, $a$ is just the axion field which
determines the torsion ($H = e^{4 \phi}\, ^*a$, in the Einstein
frame).

\vspace{3mm}

\noindent
Summarizing our results the general 4-D solution is given by
\be
\ba{l}
ds^2 = e^{4 \phi} (dt^2 + \omega_i dx^i)^2 - dx^i dx^i \qquad , \qquad
 e^{-2 \phi} = \sqrt{K F^{-1}  - |\omega_m|^2} \\
A_{\mu}^{\ r} = \frac{-1}{K - F |\omega_m|^2} \left( \ \omega^n A_{\mu}  \ |
 \ F^{-1} A_{\mu} \ \right) \quad , \quad
B_{\mu r} = \frac{1}{K - F |\omega_m|^2} \left(\   \omega_m\,A_{\mu} \
 \big| \  K\, A_{\mu}\ \right) \\
 H_{\mu\nu\rho} = - e^{2 \phi}\,\epsilon_{\mu\nu\rho\lambda}
 \partial^{\lambda} a  \quad , \quad
 e^{2 \sigma} = e^{4 ( \hat{\phi} - \phi )} =  K\, F  - F^2\,|\omega_m|^2
\quad , \quad A_{\mu} = F\, \{\,1\,,\,\omega_i\, \}
\ea
\ee
with $\omega_m = \omega^m$. Asymptotically flat solutions of (\ref{21})
and (\ref{22}) are given by
\be
K = 1 + \sum_{k=1}^{N} \frac{2 m_k}{r_k} \quad , \quad
 F^{-1} = 1 + \sum_{k=1}^{N} \frac{2 \tilde{m}_k}{r_k}
\quad , \quad \omega_m = \sum_{k=1}^{N} \frac{2 q_k^m}{r_k}
\ee
where $r_k^2 = (x-x_k)^2 + (y-y_k)^2 + (z-z_k)^2$
and the field strength $F_{ij}$ is related to the axion by
\be
F_{ij}=\partial_{[i}\omega_{j]} = \epsilon_{ijk} \partial^k a \qquad
\mbox{with} \qquad
a = \sum_{k=1}^{N} \frac{2 n_k}{r_k} \ .
\ee
For the derivation of this solution we had to make the assumption that
\mbox{$K F^{-1} - |\omega_m|^2>0$}. This was crucial for getting an
euclidean internal space. If this condition is not true the internal
metric has a positive eigenvalue (see footnote on page 4)
corresponding to one timelike internal direction and therefore a
non-compact internal space.  On the other side for the fundamental
string solution we have $K=0$, and thus, this condition is not valid.
This becomes clear if one remembers how to construct the fundamental
string solution \cite{ho/st}. One can start with a (uncharged) black
string as a direct product of the Schwarzschild metric and a flat
direction. After a Lorentz boost in the flat direction one can
dualize this direction and get a dilaton and an $H$-charge. Finally,
after performing the extremal limit one gets the fundamental string
winding around the flat direction. The coordinates $u$ and $v$ are
there light cone coordinates, which are both non compact. There are
two possibilities to avoid this problem. First is to take at least one
$\omega_m$ imaginary corresponding to a Wick rotation in the internal
coordinate with the wrong eigenvalue. This was done in \cite{be/ka/or}
to get a black hole solution from the fundamental string in 10
dimensions. Another way is to give the fundamental string first a
non-zero linear momentum and reduce it then. In this case $K\not=0$
and can be normalized to $K=1$ \cite{ho/ts1,wa}. In both cases one can
find a region with right signature.

A further remark concerns the asymptotical behavior. The upper
solution was taken to give at infinity a flat Minkowski space time.
On the other side the metric is singular where $K F^{-1} -
|\omega_m|^2=e^{-4\phi}$ becomes singular, i.e.\ for the explicit
solution (\ref{25}) at $r_k = 0$. But nevertheless we can approach
this point and the theory is still valid. There are two reasons.
First, for the chiral null model $\alpha'$ corrections are not
neglected but vanish identically, and thus, higher curvature terms
too. But there is a second may be more important reason. The vanishing
of the $\alpha'$ corrections has been shown only in an $\alpha'$
perturbative expansion. Nevertheless, there are still non-perturbative
as well as higher genus contributions to the effective action, which
have been ignored so far. But, since the string coupling constant
($g_s \sim e^{2 \phi}$) vanishes near the singularity all higher genus
and non-perturbative contributions ($\sim \exp\{- \frac{2}{g_s^2}\}$)
vanish too. Thus, the theory near the singularity is asymptotically
free (in 10 as well as in 4 dimensions), and therefore, a good play
ground for investigating the strong coupling region and space time
singularities in string theory. This feature has been discussed for
the wave or fundamental string background in \cite{ts}.

Before we investigate the general solution further let us discuss the
relation to the dilaton--axion generalization of the IWP solution
\cite{ka/ka/or/to}. The difference to that solutions is that we have
here additional gauge fields and a nontrivial modulus field. In the
case that the internal space is flat, i.e.\ $\omega_m =0$ and $ F K
=1$ both solutions coincides.  Then, we have a vanishing modulus field
($\sigma=0$), only one independent gauge field $A_{\mu}$ and from
(\ref{11}) follows that the 4-D dilaton coincides with the 10-D
dilaton. So, also the 4-D dilaton is given by a harmonic function
($-\partial^2 e^{-2 \phi} = -\partial^2 e^{-2 \hat{\phi}} =
-\partial^2 F^{-1} =0 $). If we then combine the dilaton and the axion
to one complex scalar function $\lambda = a + i e^{-2 \phi}$, this
function is harmonic too and we can write it in the form
\be
\lambda = i \left(1+ \sum_{k=1}^N \frac{2(m_k + i n_k)} {r_k}
\right) \qquad , \qquad  r_k^2 = (x-x_k)^2 + (y-y_k)^2 + (z-z_k)^2
\ee
where $m_k$ and $n_k$ are real and correspond to the masses and NUT
charges of the objects and the parameters $\vec{x}_k$ may be complex
(see \cite{ka/ka/or/to,jo/my}). For special values of the parameters
one gets other known solutions. In the case where all $\vec{x}_i$ are
real we have an extremal charged Taub-NUT solution. This metric has no
curvature singularities, but it has a conical singularity, which makes
it impossible to invert the metric along the axes $\theta=0,\pi$.
Though this singularity can be removed by making the time periodical.
Since it contains a $S_3$ topology this solution can be considered as
a cosmological model (after changing of the signature).
The other case where $n_k=0$ and $\vec{x}_k$ are complex
corresponds to introducing angular momentum. In the case of a single
source ($N=1$) this solution is a special limit of the rotating black
hole solution of Sen \cite{se} but with a ``wrong'' charge to mass ratio
(independent of the angular momentum) and thus has a naked singularity
\cite{ka/or}. Finally, if $n_k=0$ and $\vec{x}_k$ are real, i.e.\ that
$\lambda$ is pure imaginary, corresponds to a vanishing axion ($a=0$)
and field strength ($F_{ij}=0$). In this case we can diagonalize the
metric and get the extremal charged dilaton BH solution \cite{gi/ma}.
As a BH background only this solution seems to have a reasonable
interpretation. Although it has also a naked singularity, radial null
geodesic need an infinite time to reach any finite distance, and
therefore, the singularity is infinitely far away.

The question now is what are the modification if one adds the
additional gauge fields and/or adds a modulus field, i.e.\ $\omega_m
\not=0$, $ F K \not=1$. One step in this direction was already gone by
Horowitz and Tseytlin \cite{ho/ts1}. They have started with a 5-D
chiral null model and discussed the resulting 4-D theory. In our
procedure this corresponds to the case where all $\omega_m =0$ but $F
K \not=1$. As result they found one additional gauge field and a
nontrivial moduli field (compare (\ref{24})).

For the general solution (\ref{24})--(\ref{26}) the differences in
metric are contained in the dilaton $\phi$. As long as the moduli
field vanish, e.g.\ for the IWP solution, $e^{-2\phi}$ is a harmonic
function, but now it is more complicated. Let us start with the NUT
type solution, i.e.\ the functions in (\ref{25}) have a singularity in
a point.  Since the $(dt + \omega_i dx^i)$ part remains unchanged we
do not get rid of the conical singularity. The definition of
$\omega_i$ (\ref{26}) is independent of whether we have additional
gauge fields or a moduli field or not.  To get the explicit Taub-NUT
generalization we assume that we have only one source ($N=1$) and take
the spherical case ($\vec{x}_k=0$).  We can always choose our target
space coordinate system in that manner.  The harmonic function $K$ and
$F^{-1}$ can then always be written as $K = c + d F^{-1}$ and
inserting this in the Lagrangian we find that we can remove $c$ by a
translation in $v$ and can normalize $d$ ($d=1$). Thus, without any
restrictions we can set $F K =1$ or in (\ref{25}) $m=\tilde{m}$.
Inserting (\ref{25}) and a solution for $\omega_i$ \cite{ka/ka/or/to}
in (\ref{24}) we obtain for the metric ($N=1$) and the dilaton
\be
ds^2 = e^{4 \phi} \left( dt + 2 n \cos \theta d\phi \right)^2 -
\left( dr^2 + r^2 d\Omega^2 \right) \quad , \quad
e^{4 \phi} = \frac{1}{(1 + \frac{r_+}{r})(1 + \frac{r_-}{r})}
\ee
with
\be
r_{\pm} = 2(m \pm |q_m|) \ .
\ee
Here $m$ is the physical mass of the object and $q_m$ are the
electrical charges corresponding to the gauge fields $A_{\mu}^{\ m}$.
We see there is still the conical singularity at $\theta = 0, \pi$,
which can be removed by a periodic time. Furthermore, if $r_-<0$ ($m <
|q_m|$) the metric contains a pole at $r=r_-$. What happens at this
point?  If we consider the corresponding moduli field $e^{2 \sigma}$
\be
e^{2 \sigma} = \frac{(r+r_+)(r+r_-)}{(r+2m)^2} \quad \longrightarrow
 \quad \left\{
\ba{ll} 1 \quad & , \quad \mbox{ for } r \rightarrow \infty \\
    \frac{m^2 - |q_m|^2}{m^2} \quad & , \quad \mbox{ for }
    r \rightarrow 0 \ea \right.
\ee
we see that for negative $r_-$ this field has a zero and becomes
negative for $r < -r_-$. Since the moduli field is the square of
a compactification radius of an internal coordinate in this region this
direction becomes time like. In order to avoid such pathologies
we assume that $r_- \geq 0$. In addition we see, that the
compactification radius is bounded, and thus, the internal space
remains ``invisible'' for all $r$ (as long as we choose the
compactification scale sufficient small).

Secondly, let us discuss the metric for the solution with angular
momentum. In this case the functions (\ref{25}) are singular at a
circle instead a point. E.g.\ we can take the real and imaginary part
of the harmonic function $\sim (1 + \frac{2m} {x^2 + y^2 + (z + i
  \alpha)^2})$. This function is singular in the plane $z=0$ at the
circle $x^2 + y^2 = \alpha^2$. In order to get back the static
solution\footnote{In the classification of time independent metrics
static means, that one can diagonalize the metric. In contrast to
stationary metrics, which are also time independent, but the non-diagonal
part corresponds, e.g., to a non-vanishing angular momentum or Taub-NUT
charge.} ($\omega=0$) for vanishing angular momentum ($\alpha =0$)
we have to take the real part for the function (\ref{25}) and
imaginary part for the axion field (\ref{26}) (with $m=n$). If
we do that and transform the solution in the spheroidal coordinates
($x+iy = \sqrt{r^2 + \alpha^2} \sin \theta \exp\{\pm i \phi\}$; $z = r
\cos \theta$) we find
\be
\omega_{\phi} = \frac{2 m \alpha \sin^2 \theta}{R} \qquad \mbox{with}
\qquad R = \frac{r^2 + \alpha^2 \cos^2 \theta}{r}
\ee
and for the metric and the dilaton
\be
\ba{l}
ds^2 =e^{4 \phi}
\left( dt + \omega_{\phi} d\phi \right)^2 - d\vec{x}^2 \qquad , \qquad
e^{4 \phi} =  \frac{1}{(1 + \frac{r_+}{R})(1 + \frac{r_-}{R})} \\
 d\vec{x}^2 =
\frac{r^2 + \alpha^2 \cos^2 \theta}{r^2 + \alpha^2} dr^2 +
(r^2  + \alpha^2 \cos^2 \theta)d\theta^2 +(r^2 + \alpha^2)
 \sin^2\theta d\phi^2  \ .
\ea
\ee
Here, as for the Taub-NUT type solution (\ref{28}) the only influence
of the additional gauge fields and the moduli field ($r_+ \not= r_-$)
is a splitting in the dilaton field. Since the causal structure and
the singularities are not affected by this (as long as $r_- > 0$) this
metric has still a naked singularity as for $\omega_m =0$, $F K =
1$. This can also be seen by a direct comparison with a rotating black
hole solution. Starting with a 4-D Kerr solution and using the
O(d,d+p) technique Sen \cite{se} has constructed a general black hole
solution including 28 U(1) gauge fields and non-trivial moduli.  If
one performs in his metric the limit $m \rightarrow 0$, $\alpha \rightarrow
\infty$, but $m \, \sinh \alpha \, \cosh \beta = 2 M$ remains fix then
Sen's metric and dilaton becomes just (\ref{32}) (after replacing of
$|\omega_m|=M \tanh \beta$, $a = \alpha \cosh \beta$ and finally
$M\rightarrow m$; $\beta$ is a rotation parameter of O(d,d+p)).  On
the other side, Sen's solution has two horizons, which are located at
$r = m \pm \sqrt{m - a^2}$ and if we perform the upper limit we see,
that both horizons are vanish and the singularity becomes naked.

\vspace{3mm}

\noindent
To summarize, we have started with the chiral null model in 10 dimensions.
In a Kaluza--Klein approach we have reduced this model down to 4
dimensions. The field content of the 4-D theory is given by the
metric, torsion, dilaton, a modulus and 6 abelian gauge fields.  In
the limit that 5 gauge fields and the modulus field vanish our results
coincides with solutions in the literature. The effect of these
additional fields are a splitting of the dilaton pole, which is no
longer given by a harmonic function. We have argued, that the
additional gauge fields and the modulus field have no influence on the
causal structure and the singularities, and therefore, the
pathologies, e.g.\ the naked singularity for rotating black hole limit
or time periodicity in the Taub-NUT limit are still there. This is the
case at least as long as $r_- > 0$.  If $r_- =0$, i.e.\ a balance
between the mass and the electrical charges of the additional gauge
fields, the singularity in the metric is only a single pole and it
seems to be only a coordinate singularity (null geodesics reach the
singular point in a finite time).  But this question deserves further
investigation.  Also, the inclusion of Yang--Mills fields analog to
\cite{ka}, the behavior under S--duality (since the axion is harmonic
but not the dilaton this solution is not explicitly invariant under
S--duality) and cosmological interpretation of the Taub-NUT limit
could be some points of future discussions.

\vspace{5mm}

\newpage

\noindent {\large\bf Acknowledgments}
\vspace{3mm}
\newline
\noindent
I am particularly grateful to Renata Kallosh for the numerous
discussions and comments. I would like to thank Harald Dorn for
reading the manuscript and suggesting improvements. Finally, I
acknowledge the hospitality of the SLAC Theory Group during my visit
when this work was initiated.

\renewcommand{\arraystretch}{1.0}

\end{document}